\pdfoutput=1
\documentclass[prl, twocolumn]{revtex4-2}

\usepackage[T1]{fontenc}
\usepackage[utf8]{inputenc}
\usepackage{microtype, amsmath, xcolor, graphicx, hyperref, cleveref}
\hypersetup{colorlinks, 
    linkcolor={darkgray}, citecolor={darkgray}, urlcolor={darkgray},
	pdftitle={Averaged circuit eigenvalue sampling},
	pdfauthor={Steven T. Flammia}
}
\graphicspath{{figs/}}
\newcommand{\mb}{\boldsymbol}
\newcommand{\mc}{\mathcal}
\renewcommand{\ss}[1]{{\sf{#1}}}

\begin{document}

\title{Averaged circuit eigenvalue sampling}

\author{Steven T. Flammia}
\affiliation{AWS Center for Quantum Computing, Pasadena, CA 91125 USA}
\affiliation{IQIM, California Institute of Technology, Pasadena, CA 91125, USA}

\date{\today}

\begin{abstract}
We introduce ACES, a method for scalable noise metrology of quantum circuits that stands for Averaged Circuit Eigenvalue Sampling. 
It simultaneously estimates the individual error rates of all the gates in collections of quantum circuits, and can even account for space and time correlations between these gates. 
ACES strictly generalizes randomized benchmarking (RB), interleaved RB, simultaneous RB, and several other related techniques. 
However, ACES provides much more information and provably works under strictly weaker assumptions than these techniques. 
Finally, ACES is extremely scalable: we demonstrate with numerical simulations that it simultaneously and precisely estimates all the Pauli error rates on every gate and measurement in a 100 qubit quantum device using fewer than 20 relatively shallow Clifford circuits and an experimentally feasible number of samples. 
By learning the detailed gate errors for large quantum devices, ACES opens new possibilities for error mitigation, bespoke quantum error correcting codes and decoders, customized compilers, and more. 
\end{abstract}

\maketitle

Estimating errors in quantum computers is essential for progress towards fault tolerant quantum computation (FTQC)~\cite{Martinis2015}. 
An error is any undesired quantum evolution, and so errors can be as general as the set of allowed quantum dynamics, making them difficult to estimate and characterize. 
The most relevant errors in the context of FTQC can be broadly cast into the two archetypes of coherent and incoherent errors~\cite{Kueng2016}, though this is not an exclusive dichotomy. 

Coherent errors are roughly those that we wish to reduce through improved calibration or eliminate via dynamical decoupling~\cite{Viola1999}, though clever choices of quantum codes and circuits can also be tailored to handle coherent noise~\cite{Debroy2018,hu2021mitigating,zhang2021hidden}. 
These methods reach natural limits when the coherent noise becomes too complex to efficiently describe. 
While in principle coherent errors can accumulate badly during a computation~\cite{Kueng2016}, quantum error correction itself seems to reduce the coherence of noise~\cite{Huang2018,Beale2018,Iverson2020}. 

Incoherent noise, by contrast, can generally only be completely fixed by quantum error correction and fault tolerance, though near-term strategies for error mitigation could also help~\cite{Temme2017,Li2017,Endo2018,czarnik2021error,lowe2020unified}. 
Optimizing the codes, decoders, and circuits for FTQC requires a comprehensive understanding of the incoherent noise in a quantum device. 
Many techniques have been developed to estimate incoherent errors, including randomized benchmarking (RB)~\cite{Emerson2005}, interleaved RB~\cite{Magesan2012b}, simultaneous RB~\cite{Gambetta2012}, character RB~\cite{Helsen2019}, and Pauli noise estimation~\cite{Flammia2020} among others. 
Each of these techniques has in common that a general quantum noise source (which may include coherent errors) is actively \textit{averaged} to obtain an incoherent noise model with the same fidelity using randomized control techniques~\cite{Viola2005,Kern2005,Knill2005,Wallman2016,Ware2021}. 
It is this averaged noise that RB-type methods seek to estimate. 

In this paper, we show that incoherent noise, modeled as a Pauli channel, can be learned extremely efficiently using averaged circuit eigenvalue sampling, or ACES. 
It is already known that Pauli channels can be (individually) estimated efficiently and in a manner that is robust to state preparation and measurement (SPAM) errors~\cite{Harper2020,Harper2021,Flammia2020,Flammia2021}, and they are effective at modeling noise for FTQC~\cite{Geller2013,Katabarwa2015}. 
ACES goes beyond this prior work and \textit{simultaneously} estimates a large collection of Pauli noise channels associated to a quantum device. 
Indeed, we give numerical simulations showing that ACES can characterize every error rate associated to the Clifford gates in a 100 qubit quantum device using fewer than 20 circuits and a reasonable number of samples. 
Moreover, it requires only very simple classical resources to process these data, unlike other characterization techniques based on simulating or implementing general quantum circuits, or using challenging tensor network simulations~\cite{Blume-Kohout2016,Boixo2018,tao2021gleipnir,torlai2020quantum,liu2021benchmarking,cai2021sample}. 

\paragraph{The Pauli and Clifford groups.} 
The $n$-qubit \textit{Pauli group} $\ss{P}_n$ consists of $n$-fold tensor products of single-qubit Pauli operators labeled as follows. 
Let $\mb{a}$ be a $2n$-bit string $\mb{a} = a_1a_2\ldots a_{2n}$ and write $P_{\mb{a}} = i^{\mb{a}^T\Upsilon\mb{a}} \prod_{j=1}^{n} X_j^{a_{2j-1}}Z_j^{a_{2j}}$, where $X_j$ and $Z_j$ are single-qubit Paulis acting on qubit $j$, and $\Upsilon = \bigoplus_{k=1}^n \left(\begin{smallmatrix}0&1\\ 0&0\end{smallmatrix}\right)$ is such that $P_{\mb{a}}$ is always hermitian. 
The group $\ss{P}_n$ contains these $P_{\mb{a}}$, together with the overall phases $\{\pm1,\pm i\}$, composed under matrix multiplication. 
All elements of the Pauli group satisfy
\begin{align}
\label{eq:Paulicomm}
    P_{\mb{a}} P_{\mb{b}} = (-1)^{\langle \mb{a},\mb{b}\rangle} P_{\mb{b}} P_{\mb{a}}\,,
\end{align}
where the sign is determined by the binary symplectic form $\langle \mb{a},\mb{b}\rangle = \mb{a}^T (\Upsilon + \Upsilon^T)\mb{b} \bmod 2$.

The normalizer of the Pauli group inside the unitary group, modulo phases, is the \textit{Clifford group} $\ss{C}_n$, and it is generated by the controlled-NOT gate $CX_{i\to j}$ from control $i$ to target $j$, the Hadamard gate $H_j$, and the phase gate $S_j$%
~\footnote{From this definition, elements of the Clifford group are actually equivalence classes up to an overall phase, but by a slight abuse of language we can speak about a ``Clifford unitary'' to mean any representative element up to a phase and refer to uniqueness of a Clifford unitary when we really mean uniqueness up to an overall phase.}.

\textit{Pauli channels} are quantum channels of the form
\begin{align}
    \rho \mapsto \sum_{\mb{a}} p_{\mb{a}}^{\vphantom{\dag}} P_{\mb{a}}^{\vphantom{\dag}} \rho P_{\mb{a}}^\dag\,,
\end{align}
where $p_{\mb{a}}$ is a (possibly subnormalized) probability distribution called the \textit{Pauli error rates}.  
Leakage from the qubit space occurs when $\sum_{\mb{a}} p_{\mb{a}} < 1$. 
When a general quantum channel $\mc{E} = \sum_j K_j \cdot K_j^\dag$ is \textit{twirled} by the Pauli group, it becomes a Pauli channel denoted $\mc{E}^{\ss{P}}$,
\begin{align}
    \mc{E}^{\ss{P}}(\rho) = \frac{1}{4^n}\sum_{\mb{a}} P_{\mb{a}}^\dag \mc{E}(P_{\mb{a}}^{\vphantom{\dag}} \rho P_{\mb{a}}^\dag) P_{\mb{a}}^{\vphantom{\dag}}\,.
\end{align}
If $K_j = \sum_{\mb{a}} \nu_{j,\mb{a}} P_{\mb{a}}$, then the Pauli error rates of $\mc{E}^{\ss{P}}$ are $p_{\mb{a}} = \sum_j |\nu_{j,\mb{a}}|^2$. 
Thus we can speak of the Pauli error rates of a general channel by considering its Pauli twirl.
Note that we can interpret twirling as the mean of a random process where a Pauli is selected uniformly at random and applied both before and after the channel.

The eigenvectors of a Pauli channel $\mc{E}$ are just the Pauli operators. 
Indeed, from \cref{eq:Paulicomm} we have $\mc{E}(P_{\mb{b}}) = \lambda_{\mb{b}} P_{\mb{b}}$
where the \textit{Pauli eigenvalues} $\lambda_{\mb{b}}$ are,
\begin{align}
    \lambda_{\mb{b}} = \sum_{\mb{a}} (-1)^{\langle \mb{a},\mb{b}\rangle} p_{\mb{a}}\,.
\end{align}
This equation can be inverted to express the error rates in terms of the eigenvalues%
~\footnote{Note that $\lambda_{\mb{a}}$ and $p_{\mb{a}}$ are essentially Fourier transform pairs since the transform relating them is the Walsh-Hadamard transform (up to a permutation). 
A helpful intuition is that $\lambda$ lives in the ``time domain'' where we can efficiently sample, and $p$ lives in the ``frequency domain'' where we wish to reconstruct the signal.},
\begin{align}
\label{eq:p}
    p_{\mb{a}} = \frac{1}{2^n} \sum_{\mb{b}} (-1)^{\langle \mb{a},\mb{b}\rangle} \lambda_{\mb{b}}\,.
\end{align}

We now introduce a ``$\mc{G}$-twisted'' Pauli twirl. 
For a given Clifford $\mc{G}$ and Pauli $P_{\mb{a}}$, let $P_{\mb{a}'} = \mc{G}(P_{\mb{a}})$. 
Note that since $\mc{G}$ is unitary, the set of all $\mb{a}$ and $\mb{a}'$ are in one-to-one correspondence. 
We wish to expand a noisy gate as $\widetilde{\mc{G}} = \mc{G}\mc{E}$ for some general noise channel $\mc{E} = \mc{G}^\dag\widetilde{\mc{G}}$. 
Intuitively, $\mc{E}$ is close to the identity, though the definition doesn't assume that. 
Then the $\mc{G}$-twisted twirl of $\widetilde{\mc{G}}$ is
\begin{align}
\label{eq:Gtwistedtwirl}
    \widetilde{\mc{G}}^{\mc{G}\ss{P}}(\rho) = \frac{1}{4^n}\sum_{\mb{a}} P_{\mb{a}'}^\dag \widetilde{\mc{G}}(P_{\mb{a}}^{\vphantom{\dag}} \rho P_{\mb{a}}^\dag) P_{\mb{a}'}^{\vphantom{\dag}} = \mc{G}\bigl(\mc{E}^{\ss{P}}(\rho)\bigr)\,.
\end{align}
From the last equality, we see that the $\mc{G}$-twisted twirl isolates the Pauli noise around a given noisy implementation $\widetilde{\mc{G}}$ of an ideal Clifford gate $\mc{G}$. 

$\mc{G}$-twisted twirled channels have an analogous eigendecomposition to a Pauli twirled channel, but with the notion of \textit{generalized eigenvector}. 
Given such a channel $\widetilde{\mc{G}}^{\mc{G}\ss{P}}$, the generalized eigenvectors with respect to $\mc{G}_0$ are vectors such that $\widetilde{\mc{G}}^{\mc{G}\ss{P}}(v) = \lambda \mc{G}_0(v)$. 
We see from \cref{eq:Gtwistedtwirl} that choosing $\mc{G}_0 = \mc{G}$ gives exactly the Paulis as the generalized eigenvectors with eigenvalues given by the Pauli eigenvalues of the noise map $\mc{E}^{\ss{P}}$.

\paragraph{Averaged circuits.}
Let us consider a Clifford circuit (i.e., a circuit composed solely of $CX$, $H$, and $S$ gates or an equivalent generating set), denoted $\mc{C}$. 
Any physical implementation of these circuits will be noisy, and we seek to characterize the incoherent Pauli-averaged noise in these circuits, specifically in the generators used to create the circuits. 
To that end, from the circuit $\mc{C}$ we create a new ensemble of circuits $\mc{C}^{\ss{P}}$ by sampling a $\mc{G}$-twisted Pauli twirl across each Clifford circuit element and recompiling the Pauli gate. 
This approach to Pauli frame randomization is known as randomized compiling~\cite{Wallman2016}. 
Each circuit in the ensemble implements the same unitary operation, but now the noise has been averaged over the Pauli group. 
In Ref.~\cite{Wallman2016}, it was proven that circuits sampled in this way yield on average a circuit that interleaves Pauli-averaged noise with ideal gates (except possibly in the final measurement step). 
This result rigorously holds whenever the noise on each Pauli gate is the same, and furthermore Ref.~\cite{Wallman2016} provides some robustness guarantees in the event that this assumption is perturbatively violated%
~\footnote{While the gate-independent noise assumption may seem unrealistic, it should be noted that the successful and widely used method of standard RB makes the \textit{much} stronger assumption that the noise is gate-independent across all $n$-qubit Clifford gates, whereas ACES weakens this substantially to just the Pauli gates.}. 

These considerations motivate considering only \textit{averaged circuits}, denoted $\mc{C}^{\ss{P}}$, so that the noisy physical implementations will have the form 
\begin{align}
\label{eq:stdform}
    \widetilde{\mc{C}}^{\ss{P}} = \widetilde{\mc{G}_T}^{\mc{G}_T\ss{P}}\ldots\widetilde{\mc{G}_1}^{\mc{G}_1\ss{P}} = \mc{G}_T\mc{E}_{\mc{G}_T} \ldots \mc{G}_1\mc{E}_{\mc{G}_1}\,. 
\end{align}

\paragraph{Eigenvalue sampling.}
Let us suppose for the moment that a given circuit $\mc{C}$ ideally implements the identity unitary. 
Under the gate-independent noise assumption, it follows that the noisy implementation of the averaged circuit, $\widetilde{\mc{C}}^{\ss{P}}$, will be a Pauli channel. 
It therefore has Pauli eigenvalues, namely $\widetilde{\mc{C}}^{\ss{P}}(P_{\mb{a}}) = \Lambda_{\mc{C},\mb{a}} P_{\mb{a}}$, where we use capital $\Lambda$ to denote this circuit-level eigenvalue. 
Because of the gate-independent noise assumption, this eigenvalue depends only on the eigenvector ($P_{\mb{a}}$) and on the circuit ($\mc{C}$), so it is labeled accordingly as $\Lambda_{\mc{C},\mb{a}}$. 

If the circuit $\mc{C}$ does not implement the identity unitary, but rather some net Clifford operation, something similar still holds. 
If the ideal circuit maps an input Pauli $P_{\mb{a}}$ to an output Pauli $\mc{C}(P_{\mb{a}}) = \pm P_{\mb{a}'}$, then the overall $\pm$~sign and the value of $\mb{a}'$ can be efficiently computed~\cite{Aaronson2004}. 
The noisy version of the circuit will give an averaged operator that satisfies the generalized eigenvalue equation
\begin{align}
\label{eq:eigeq}
\widetilde{\mc{C}}^{\ss{P}}(P_{\mb{a}}) = \Lambda_{\mc{C},\mb{a}} \mc{C}(P_{\mb{a}}) = \pm \Lambda_{\mc{C},\mb{a}} P_{\mb{a}'}\,.
\end{align}
From the orthogonality of the Pauli basis, it follows that
\begin{align}
\label{eq:Lambda}
    \Lambda_{\mc{C},\mb{a}} = \pm\frac{1}{2^n} \mathrm{Tr}\bigl(P_{\mb{a}'}  \widetilde{\mc{C}}^{\ss{P}}(P_{\mb{a}})\bigr)\,,
\end{align} and this suggests a prescription for estimating the (generalized) eigenvalue $\Lambda_{\mc{C},\mb{a}}$ that we call eigenvalue sampling. 

To estimate $\Lambda_{\mc{C},\mb{a}}$ via eigenvalue sampling, let us focus on the case where $P_{\mb{a}}$ is a single-qubit Pauli. 
We begin by selecting uniformly at random an eigenstate $\psi_{\pm}$ on the support of $P_{\mb{a}}$ having eigenvalue $\pm1$ (ignoring the other registers). 
Then we send $\psi_{\pm}$ into a randomly chosen element in the circuit ensemble $\mc{C}^{\ss{P}}$ and measure the output in the basis defined by $P_{\mb{a}'}$.
Our overall estimate for $\Lambda_{\mc{C},\mb{a}}$ consists of measuring $N$ independent experiments and taking the difference of the sample averages between the $\psi_+$ and $\psi_-$ experiments. 
It is easy to check that this differencing trick makes \cref{eq:Lambda} hold in expectation, so this is an unbiased estimator of $\Lambda_{\mc{C},\mb{a}}$. 
This sampling strategy was first analyzed in Ref.~\cite{Flammia2011}, and it is straightforward to generalize to the $n$-qubit case. 
Note that it will be most efficient if the support of $P_{\mb{a}}$ and $P_{\mb{a}'}$ are relatively small, and also that Paulis with disjoint support can implement such measurements simultaneously.

\paragraph{Relating circuit and gate eigenvalues.}
We have seen how eigenvalue sampling on averaged circuits gives us access to the (generalized) Pauli eigenvalues $\Lambda_{\mc{C},\mb{a}}$ in the implemented circuit ensemble $\widetilde{\mc{C}}^{\ss{P}}$. 
This is already a useful method for estimating the quality of the circuit implementation $\widetilde{\mc{C}}$, since it can be interpreted as a fidelity-like measure for how faithfully the circuit executes given the input $P_{\mb{a}}$. 
However, we seek to characterize not just the global circuit noise, but the error rates associated to the constituent gates as well. 
How do the (generalized) eigenvalues of the individual gates relate to the eigenvalue of the total circuit $\mc{C} = \mc{G}_T \ldots \mc{G}_1$?

Let us apply the generalized eigenvalue relation sequentially to the gates in a Clifford circuit. 
For the first gate we obtain $\widetilde{\mc{G}}^{\ss{P}}_1(P_{\mb{a}_1}) = \lambda_{1,\mb{a}_1} \mc{G}_1(P_{\mb{a}_1}) = (\pm)_1 \lambda_{1,\mb{a}_1} P_{\mb{a}_2}$. 
Acting on this with $\widetilde{\mc{G}}^{\ss{P}}_2$, we obtain 
\begin{align*}
    \widetilde{\mc{G}}^{\ss{P}}_2\widetilde{\mc{G}}^{\ss{P}}_1(P_{\mb{a}_1}) & = (\pm)_1 (\pm)_2 \lambda_{1,\mb{a}_1} \lambda_{2,\mb{a}_2} \mc{G}^{\ss{P}}_2(P_{\mb{a}_2}) \\
    & = (\pm)_1 (\pm)_2 \lambda_{1,\mb{a}_1} \lambda_{2,\mb{a}_2} P_{\mb{a}_3}\,.
\end{align*}
Continuing in this fashion, we find that 
\begin{align}
    \widetilde{\mc{C}}^{\ss{P}}(P_{\mb{a}_1}) = (\pm) \prod_{k=1}^{T} \lambda_{k,\mb{a}_k} \mc{C}(P_{\mb{a}_1})\,,
\end{align}
where the overall sign and the individual $\mb{a}_k$ can be computed efficiently~\cite{Aaronson2004}. 
Comparing with \cref{eq:eigeq}, we see that $\Lambda_{\mc{C},\mb{a}_1} = (\pm)\prod_k \lambda_{k,\mb{a}_k}$. 
We will use the freedom to reinterpret the sign of the input Pauli $P_{\mb{a}_1}$ to ensure that we always have a $+$ sign in this equation, and therefore we have the relation
\begin{align}
\label{eq:prod}
    \Lambda_{\mc{C},\mb{a}_1} = \prod_{k=1}^{T} \lambda_{k,\mb{a}_k}\,.
\end{align}
With this sign convention, in the regime of interest $\Lambda_{\mc{C},\mb{a}_1}$ is positive and not too small. 
We therefore focus on sets of circuits $\mc{C}_k$ and input labels $\mb{a}_{k_i}$ such that $\Lambda_{\mc{C}_k,\mb{a}_{k_i}}$ is always larger than, say, $1/2$, and gates where $\lambda_{k,\mb{a}_k} > 0$.

\paragraph{Estimating gate errors via ACES.} 
We now consider a circuit $\mc{C}_k$ and an input label $\mb{a}_{k_i}$; we give this combination a composite index $\mu = (\mc{C}_k,\mb{a}_{k_i})$, where $\mu = 1,\ldots M$. 
From the above discussion, we can obtain an empirical estimate $\hat{\Lambda}_\mu$ of $\Lambda_\mu$ by eigenvalue sampling on the averaged circuit ensemble for the circuit/input label $\mu$. 
Similarly, we assemble all gate-level eigenvalues under a single index to get $\lambda_\nu$, where $\nu$ labels pairs of gates and Paulis whose noise we wish to model, with $N$ total model parameters. 
Since all eigenvalue quantities are positive in the regime of interest, we can introduce new variables, 
\begin{align}
\label{eq:exp}
    \Lambda_\mu = \mathrm{e}^{-b_\mu}\,, \quad \lambda_\nu = \mathrm{e}^{-x_\nu}\,. 
\end{align}
The new variables are related by the \textit{linear} equations
\begin{align}
\label{eq:Ax=b}
    \sum_\nu A_{\mu\nu} x_\nu = b_\mu\,.
\end{align}
We refer to the $M\times N$ matrix $A$ as the \textit{design matrix}. 
Once enough independent equations are obtained so that $A$ has rank $N$, an estimate for $\mb{x}$ can be obtained in any number of ways~\footnote{Although a simple least-squares fit is a consistent estimator, more sophisticated estimators would take into consideration the constraint that the channel eigenvalues arise from physical maps. 
The truncated least-squares estimator used in the numerics is a step towards such an estimator, but we leave the exploration of optimal estimators for future work.}%
, most straightforwardly via least squares as $\hat{\mb{x}} = A^+\hat{\mb{b}}$, where $\hat{\mb{b}}$ denotes an empirical estimate for $\mb{b}$ and $A^+$ is the pseudoinverse of $A$. 
Inverting \cref{eq:exp} subsequently gives us estimates for $\lambda_\nu$, and Pauli error rates can be obtained by using \cref{eq:p}. 

The precision of our estimate depends in part on the choice of $A$, as well as the precision of the initial estimates of the $\Lambda_\mu$. 
The estimates for $\lambda_\nu$ are always accurate in the sense that these are consistent estimators, however they will in general have some bias. 
In the numerical simulations below, no attempt was made to find optimal designs $A$, and only random choices were used. 
We leave open the question of finding optimal design matrices that maximize the precision and accuracy of these estimators. 

\paragraph{Correlations and SPAM.}
The ACES methodology is flexible enough to allow independent estimation of SPAM errors as well as space- and/or certain time-correlated errors. 
To estimate measurement noise, we simply add a list of variables $x_\nu$ associated to each Pauli measurement error that we wish to model. 
We caution that separating preparation errors from measurement errors will not be possible if they are introduced into the model in a symmetric way (because then $A$ will not have rank $N$); this problem is not unique to ACES however~\cite{Blume-Kohout2016} and we do not attempt to resolve it here. 

To handle space-correlated errors, we reinterpret the gates that generate our circuits to come in correlated groups. 
For example, if we want to model correlated noise between the Hadamard gates $H_1$ and $H_2$, we could have separate variables for the gates $H_1$, $H_2$, and $H_1\otimes H_2$. 
This is analogous to interleaved~\cite{Magesan2012b} and simultaneous RB~\cite{Gambetta2012}, except that all of the data are used to fit all of the gates and correlations symmetrically and simultaneously. 

Limited forms of time-correlated errors can be handled similarly by introducing variables for pairs of gates in time. 
For example, if the noise on $H_1$ depends on whether $S_1$ was applied or not right before, then we can introduce separate variables for these cases. 

The only condition for a unique and consistent estimate in all of these scenarios is that the design matrix $A$ has rank $N$. 
If $A$ were random, then we only need as many equations as unknowns for this to hold with high probability. 
From this heuristic, we expect that the number of experiments should be about as large as, or a little larger than, the number of variables.

\begin{figure*}[t!]
    \centering
    a) \includegraphics[scale=0.5]{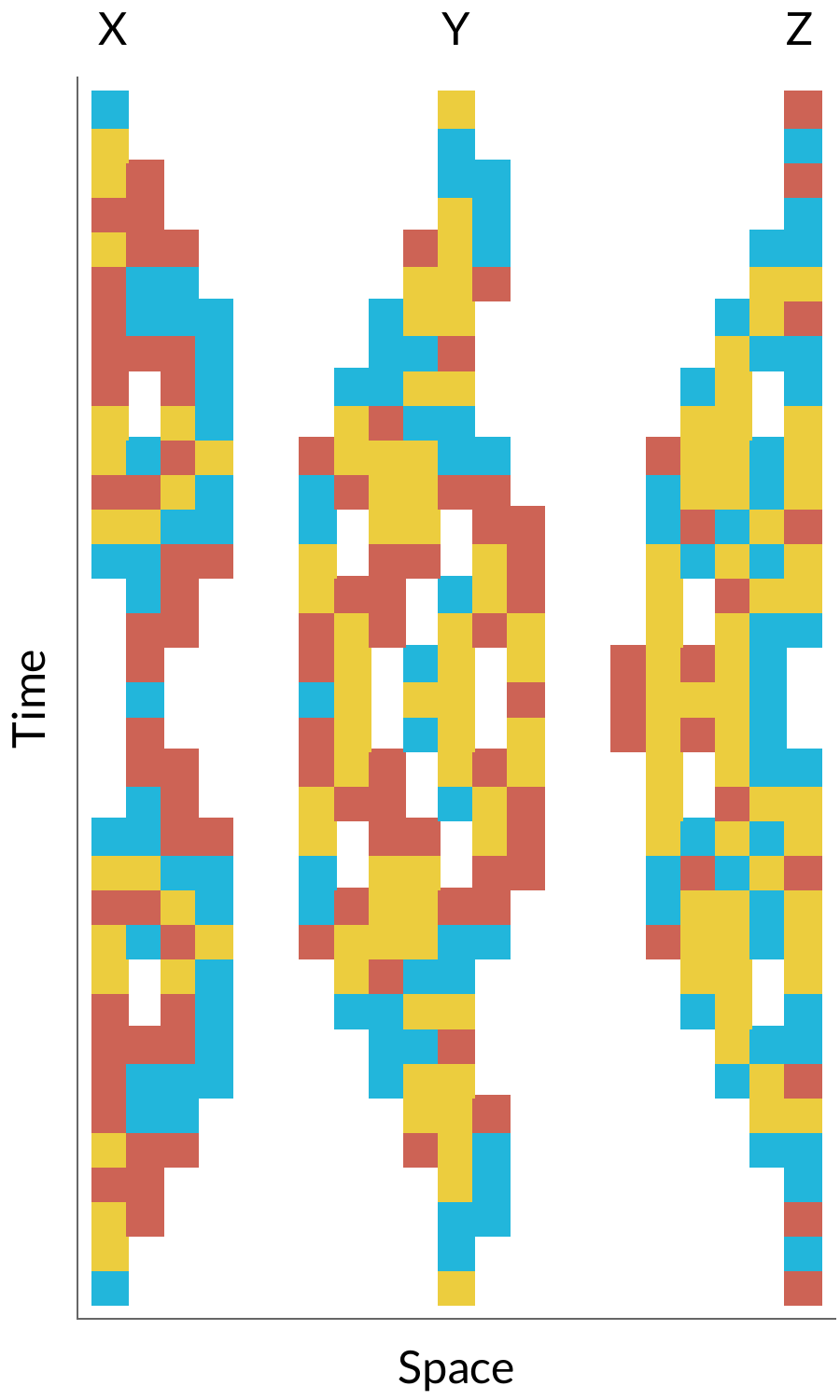}
    \quad 
    b) \includegraphics[scale=0.5]{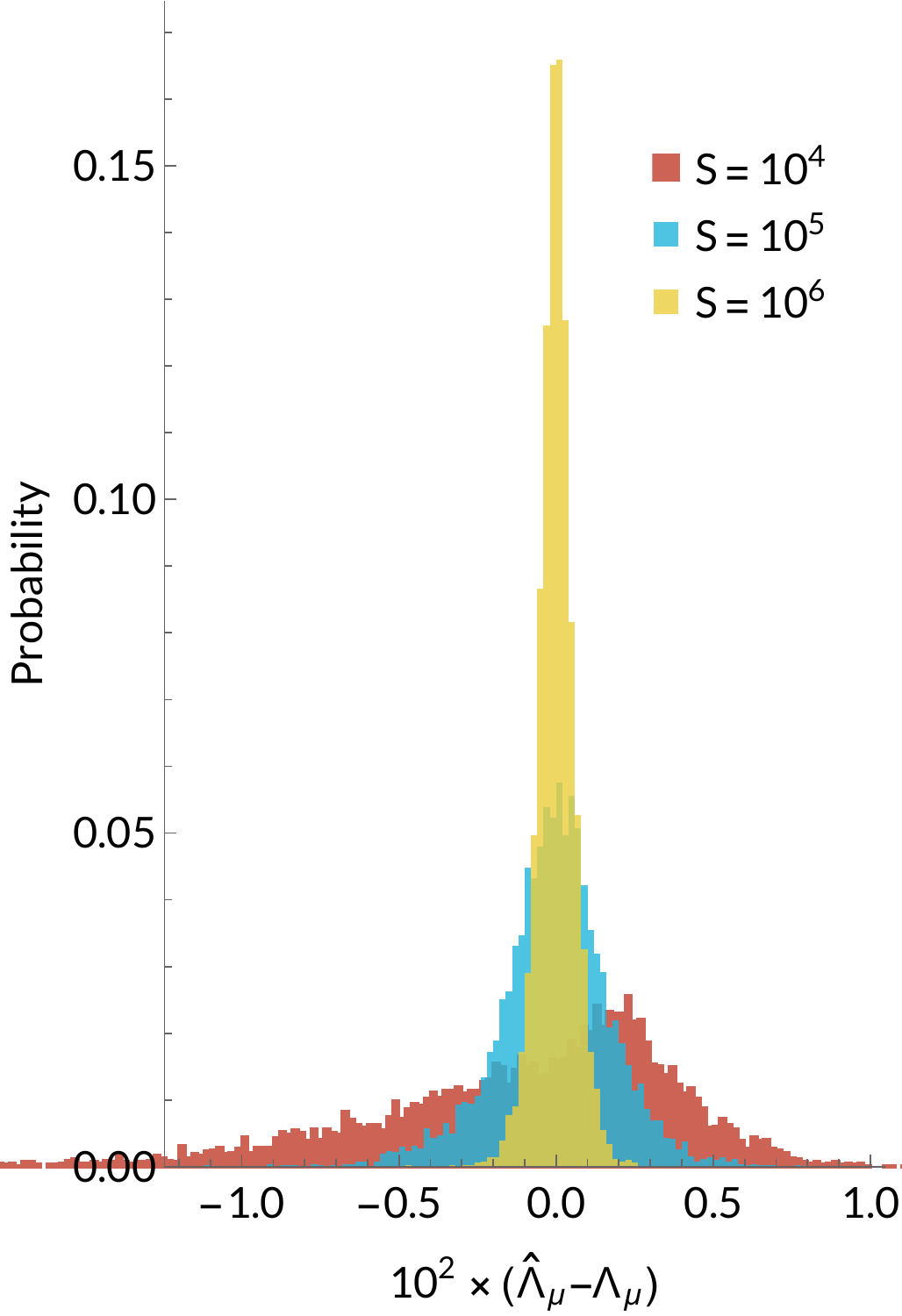}
    \quad
    c) \includegraphics[scale=0.5]{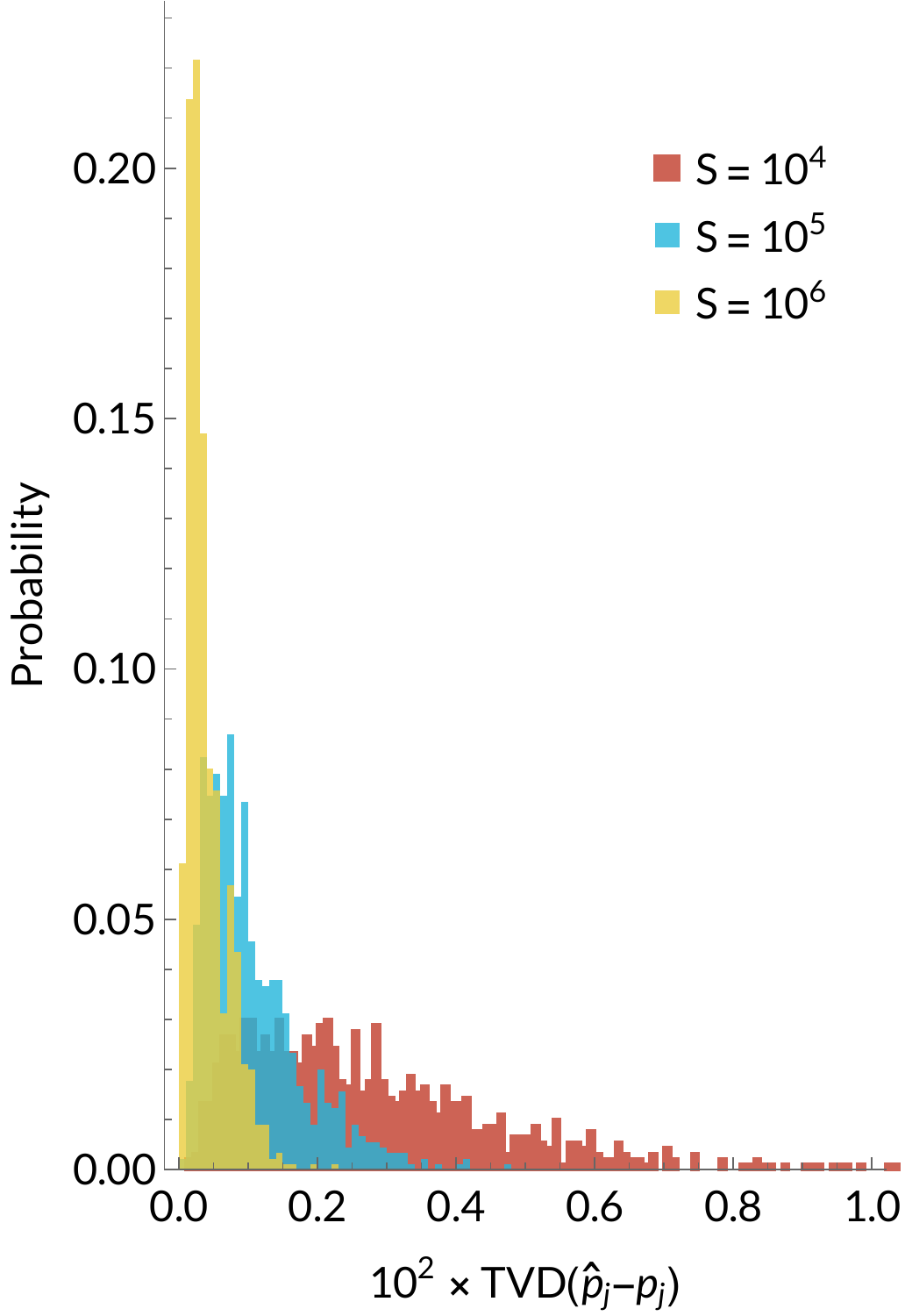}
    \caption{a) Sending $X$, $Y$, and $Z$ Paulis (blue, yellow, and red, resp.) through a small ``mirror circuit'' (i.e., one of the form $UU^\dag$) with $n=21$ qubits, depth $d=34$, and nearest-neighbor gates in 1D. 
    Normalized histograms of b) the absolute error for the $\mu$th estimated circuit eigenvalue $\hat{\Lambda}_\mu$, and c) the total variation distance (TVD) for the estimated Pauli error rates $\hat{p}_j$ of the noisy gate $\tilde{\mc{G}}_j$ in a $n=100$ qubit simulation. 
    There are $898$ gates (including measurements) in the model, and $10,155$ estimated circuit fidelities (which are estimated in large batches due to the $n$-bit measurements) to estimate $N = 5070$ parameters.
    Plots are for a number of samples $S$ per $\Lambda_\mu$ of $10^4, 10^5$, and $10^6$.}
    \label{fig:fig}
\end{figure*}

\paragraph{Numerical results.}
We now demonstrate the scalability of ACES via numerical simulations. 
Rigorous proofs of the consistency of ACES and bounds on the sample complexity will be presented elsewhere. 

We consider the most general model of inhomogeneous but uncorrelated noise, plus readout errors%
~\footnote{We do not model state preparation errors, only readout errors, for the reason discussed in the main text that these are not separately identifiable.}. 
In this model on $n$ qubits there are $O(n)$ variables: $CX$ gates acting between neighbors, together with six single-qubit Clifford gates (modulo the Paulis), and independent readout errors on each qubit in each Pauli basis. 

We generated $C = 19$ random 1D Clifford circuits on $n=100$ qubits of varying depths from $d=2$ up to $d=89$. 
The sum of all the circuit depths, including the measurement rounds, was $354$. 
We then computed the circuit eigenvalues obtained from sending in all single-qubit Paulis and, on some circuits, two-qubit Paulis on nearest neighbors as well. 
We found it challenging to generate a rank-$N$ design matrix $A$ using the ``mirror circuits'' shown in \Cref{fig:fig}a, so we padded each mirror circuit with a depth 5 random circuit layer at the end. 
This means that the Paulis measured at the output had, in some cases, weight as high as 6, though most still had weight 1 or 2. 
Constant-weight Pauli operators can nonetheless be estimated efficiently from single-qubit Pauli measurements~\cite{evans2019scalable,Huang2020,huang2021efficient}, and this only increases the sample complexity by a constant factor. 
We then generated a ``true'' noise model by assigning to each gate random Pauli error rates consistent with the estimates reported in the Arute \textit{et al}. experiment~\cite{Arute2019}. 
The entire implementation can be found in the associated Mathematica notebook accompanying this manuscript~\cite{ACESgithub}.

Despite its seeming simplicity, this model still has $N = 5070$ parameters. 
Even under the simplifying assumptions of RB with Clifford averaging where the noise is depolarizing on each gate, there would still be $798$ parameters (neglecting SPAM) to be estimated through interleaved RB, and even then the required Clifford randomizations would be prohibitively expensive. 

ACES estimates all of these parameters with just these 19 random circuits (and their Pauli randomizations). 
This is possible because each measurement is an $n$-bit measurement, so many parameters are estimated in parallel. 

In Figs.~\ref{fig:fig}b and \ref{fig:fig}c we plot the convergence of the ACES estimate as a function of $S$, the number of samples per circuit eigenvalue estimate.
Estimates $\hat{x}_\nu$ of the model parameters $x_\nu$ were obtained from the simulated data by solving \cref{eq:Ax=b} with the simplest possible estimator, a truncated least-squares estimate (i.e., finding the least squares solution and truncating any negative values). 

Counting an $n$-bit measurement as one sample, the total sample complexity is $O(S C)$ where $C$ is the number of different averaged circuits used, in this case $C = 19$.
Results are shown for $S=10^4, 10^5, 10^6$. 
Even for $S=10^4$, nearly all circuit eigenvalue estimates (\ref{fig:fig}b) are within 1\% of the true answer, and the total variation distance (TVD) between the estimated and true Pauli error rates on each gate are within .64\% with 95\% confidence. 
This latter figure improves to .1\% with 95\% confidence for $S=10^6$, a remarkably precise estimate given how many parameters there are and that no regularization was used to avoid potential overfitting.

\paragraph{Discussion.}
There are many potential applications for ACES, and many avenues for improving and extending it as well.
For example, in addition to the tailored codes and decoders mentioned above, error mitigation is one of the most natural applications of ACES~\cite{Temme2017,Li2017,Endo2018,czarnik2021error,lowe2020unified}, and it can also be used to debias estimates of classical shadows following the ideas in Refs.~\cite{chen2020robust,koh2020classical,berg2021modelfree}. 
Regarding extensions, while we have focused entirely on Clifford gates, it is easy to see that ACES can accommodate circuits with a constant number of $T$ gates in specific configurations. 
However, extending beyond this to universal gate sets in general is an important question for future research. 
A differential analysis suggests that obtaining circuit eigenvalue estimates such that $\hat{\Lambda}_\mu = \Lambda_\mu \pm \epsilon \Lambda_\mu$ suffices to obtain gate-level eigenvalue estimates of order $\hat{\lambda}_\nu = \lambda_\nu \pm O\bigl(\|A^+\| \epsilon\bigr)\lambda_\nu$. 
Thus, finding circuits, Pauli inputs, and noise models whose associated design matrix minimizes $\|A^+\|$ could help optimize the sample efficiency of ACES. 
There are additional desiderata for the design matrix, such as requiring only few experiments and using circuits that map few-qubit Paulis to few-qubit Paulis. 
Finding a general understanding of which circuits behave best is an open question. 
While ACES can test for correlations in a given noise model, it would be more powerful to include a large model and then search for dominant correlations by enforcing sparsity. 
One way forward might be to test clusters of gates for inclusion using methods such as group LASSO~\cite{Yuan2006}. 
Finally, the most obvious open problem is to implement ACES in a near-term experiment. 

We thank Laura DeLorenzo, Robin Harper, Robert Huang, Alex Kubica, Ryan O'Donnell, Colm Ryan, Prasahnt Sivarajah, and Giacomo Torlai for discussions and the ARO QCISS program grant W911NF2110001 for support. 

\bibliography{refs}
\end{document}